\documentclass[fleqn,twoside]{article}
\usepackage{espcrc2}
\usepackage{amsmath,amssymb}
\usepackage{ae,aecompl,cancel}
\usepackage{yfonts}

\usepackage{graphicx}
\usepackage[figuresright]{rotating}

\newcommand{\be}{\begin{equation}}
\newcommand{\ee}{\end{equation}}
\newcommand{\beqa}{\begin{eqnarray}}
\newcommand{\eeqa}{\end{eqnarray}}

\def\dA{{\cal A}}

\newcommand{\Dis}{{\tilde{d}}}
\newcommand{\Diff}{D}

\title{On the Einstein Relation in holographic systems at finite baryon density}

\author{Javier Mas\address{Departamento de F\'\i sica de Part\'\i culas,
Universidade de Santiago de Compostela and
Instituto Galego de F\'\i sica de Altas Enerx\'\i as (IGFAE),E-15782, Santiago
de Compostela, Spain}, Jonathan P. Shock, Javier Tarr\'\i o}

\begin{document}

\begin{abstract}
We study the conductivity, susceptibility and diffusion of a strongly coupled quark gluon plasma from the holographic perspective. We calculate general expressions for these quantities in the presence of finite baryon density and show that in this context, for the D3/D7 intersection the Einstein relation holds, providing another non-trivial check of the holographic correspondence at finite temperature.
\vspace{1pc}
\end{abstract}

\maketitle
\setcounter{footnote}{0}

\section{Introduction}
In this note we consider the validity of the Einstein relation in the context of strongly coupled plasmas with finite chemical potential. The Einstein relation in question relates two transport coefficients, the diffusion constant, $D$, and the conductivity, $\sigma$, with an equilibrium quantity, the susceptibility, $\chi$. We illustrate here that there is a correct prescription for the calculation of these three quantities within the context of holographic flavors in the quenched approximation for which Einstein's relation holds. Note that this has been studied in the absence of baryon number previously, for instance in \cite{Kovtun:2008kx}, while the universality of transport coefficients has been discussed extensively in \cite{Iqbal:2008by}. Before coming onto this relation we set up the context of our calculation.

We consider a holographic construction, with $N_f$ Dq branes embedded in the metric of $N_c\gg N_f$ Dp branes in the quenched approximation. The probe Dq branes wrap an $n$-sphere and the space transverse to them is spanned by $\psi\in [0,1)$ and angular coordinates $\varphi^i$. The  embedding of the probe brane and the $U(1)$ gauge field strength are parametrized by functions $\psi(r)$ and $A_\mu(r)$, whose equations of motion are derived  from the DBI action. We are interested in solutions that involve a non-zero value for  $A_t$, dual to a non-zero chemical potential in the gauge theory. The charge density $n_q$ is a constant of motion and can be obtained
from the electric displacement  \cite{Kobayashi:2006sb}
\be
n_q = \frac{\delta {\cal S}_{DBI}}{\delta A_t'(r)} \, .\label{betadis}
\ee
We shall here deal with the computation of retarded correlators and consider fluctuations in the worldvolume fields of the following form
\beqa
\hspace{-0.5cm}\psi(r,x)& \rightarrow &\psi(r) + \epsilon e^{-i(\omega x^0 - q x^1)} \Psi(r)\, , \nonumber\\
\hspace{-0.5cm}A_\mu(r,x) &\rightarrow &A_\mu(r) + \epsilon e^{-i(\omega x^0 - q x^1)}\dA_\mu (r) .
\eeqa
The linearized equations for the perturbations $\Psi$ and $\dA$ are derived by expanding the DBI lagrangian  up to  ${\cal O}(\epsilon^2)$.
A detailed analysis of these equations can be found in \cite{Mas:2008jz,Mas:2008qs}.
\section{The Conductivity}
The conductivity is obtained from the zero-frequency slope of the trace of the spectral function \cite{CaronHuot:2006te,Mateos:2007yp}
\be
\sigma = 
\frac{1}{2p} \lim_{\omega\to 0}\frac{\chi^\mu{_\mu}(\omega, {\bf q} = 0)}{\omega} \, .
\label{condchi}
\ee
The spectral function is defined in terms of the two point Greens function of electromagnetic currents and in the limit of zero frequency only the transverse part of the Greens function contributes,  given in terms of the boundary DBI action as \cite{Son:2002sd}
\be
 \Pi_{\small\perp}(k)=\left.
 {\mathcal N} e^{-\phi} \sqrt{-\gamma} \gamma^{ii} \gamma^{rr} \frac{ E'_{\perp}(r) E_{\perp}(r) }{|E_\perp(r_{B})|^2}
 \right\vert_{r_{B}}.
 \label{traspect}
\ee
where $E_\perp=\omega  {\cal A}_{2,3}$, $\gamma^{ab}$ are inverse components of $\gamma_{ab}=g_{ab}+2\pi\alpha' A_{ab}$, $g_{ab}$ is the pullback metric on the probe brane and ${\cal N}$ is a constant factor. By  expanding the equation of motion of $E_\perp$ to linear order in $\omega$ and $q$ and imposing regularity on the horizon, we obtain
\be
\sigma ={\mathcal N}\, e^{-\phi} \sqrt{\gamma \gamma^{00} \gamma^{rr}} \gamma^{ii}\Big|_{r_H}\, .
\label{sigma}
\ee
\section{The susceptibility \label{secsusc}}
The susceptibility is an equilibrium quantity, which can be defined holographically as
\be
\chi =  \left(\partial_{n_q}\mu\right)^{-1}|_T=\left(\int_{r_H}^{r_B}\frac{d A'_t(r) }{d  n_q}dr 
\right)^{-1} \, .
\label{chidef}
\ee
From (\ref{betadis}) we can relate $A'_t(r)$  to the charge density
\be
A_t' =  \frac{n_q}{{\cal N}}	\sqrt{ \frac{-\gamma_{00}\gamma_{rr}}{\displaystyle e^{-2\phi} \gamma_{ii}^p \gamma_{\theta\theta}^n + 
\displaystyle (2\pi\alpha')^2\frac{n_q^2}{{\cal N}^2}}} ~ . 
\label{mudenq}
\ee

Notice that there is an implicit $n_q$ dependence
 in the brane embedding $\psi(r)$. After some algebra we can use (\ref{chidef}) and (\ref{mudenq}) to obtain:
\be
\hspace{-0.03cm}\chi ={\cal N}\left( \int_{r_H}^{r_B}
\rule{0mm}{6mm}
\frac{1+ n_q \left(  \Xi\, \psi_{,n_q}+ \Delta\, \psi'_{,n_q}
\right)}{e^{-\phi}  \sqrt{-\gamma}\gamma^{00}\gamma^{rr} }
\right)^{-1},
\label{suscepbaryon}
\ee 
with
$\Delta,\Xi = \Delta,\Xi\left(\gamma^{ab}(r),\psi(r)\right)$ (see \cite{Mas:2008qs} for the complete expressions).

\section{Charge diffusion \label{secdifu} at finite baryon density}
In  \cite{Kovtun:2003wp,Starinets:2008fb} the diffusion of a generic conserved charge was studied using the membrane paradigm and the AdS/CFT correspondence. Here we derive the charge diffusion constant from the pole in the longitudinal correlator at finite baryon density and quark mass.

In the presence of a background value for $A_t$, the gauge invariant longitudinal perturbations, $E_{||}= q\dA_{t}+\omega \dA_{1}$, mix with the scalar fluctuations  of the brane profile $\Psi(r)$. The coupled second order equations of motion can be found in reference   \cite{Mas:2008jz}.
All of the coefficients of $\Psi$ are of order $q$ implying that the natural variable
is $\tilde \Psi = q \Psi$.
In the case of the diffusion pole, we are looking for a dispersion relation of the form $\omega = - i \Diff q^2 + ...$ thus the natural scaling is given in terms of a variable $\lambda$ as follows $\omega \to \lambda^2 \omega, \, q\to \lambda q$. Expanding to order $\lambda^0$ the coupled equations for the perturbations and integrating we obtain
\be
E_{||}^{(0)}(r) =1+ \int  \frac{ - i C\displaystyle\frac{q^2}{\omega}+   n_q \frac{2\pi \alpha'}{\cal N} (\Delta \tilde \Psi_{(0)}' + \Xi \tilde \Psi_{(0)})}{\sqrt{-\gamma}\gamma^{00}\gamma^{rr}}\, .
\label{eparallel}
\ee
where the constants of integration are fixed by studying the Frobenius expansion around the horizon and $\tilde{\Psi}_{(0)}$ is the zeroth order in the expansion of $\tilde{\Psi}$ in $\lambda$. From the boundary condition $E_{||}^{(0)}(r_B)=0$ we solve for the zero in this mode and by comparing with $\omega=-i Dq^2+{\cal O}(q^3)$ obtain
\be
\hspace{-0.4cm}\Diff = \frac{ \left.e^{-\phi} \sqrt{\gamma \gamma^{00} \gamma^{rr}} \gamma^{ii}\right\vert_{ r_H} \int_{r_H}^{r_B} \frac{dr}{e^{-\phi} \sqrt{-\gamma}\gamma^{00}\gamma^{rr}}}{1+\displaystyle n_q \frac{2\pi \alpha'}{{\cal N}}  \int  \frac{  (\Delta\tilde \Psi_{(0)}' + \Xi \tilde \Psi_{(0)})}{\sqrt{-\gamma}\gamma^{00}\gamma^{rr}}}.
\label{diffbaryon}
\ee
Note that in contrast to the conductivity and susceptibility this expression involves fluctuations.

\section{Numerics, the D3/D7 case study \label{secd3d7}}
Here we focus on the case of a D3/D7 flavour setup. For finite baryon density, only results concerning the conductivity have been reliably established until now and it can be easily shown that our expression for $\sigma$ matches with the expression in \cite{Karch:2007pd}. 
\begin{figure}[ht]
\begin{center}
\includegraphics[scale=0.8]{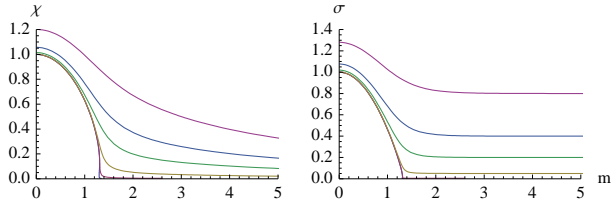}
\caption{\em \label{twoplots}
$\chi$ (in units of $N_cN_f T/4\pi$), and  $\sigma$ (in units of $N_c N_f T^2/2$), against the quark mass defined by the D7 embedding. From bottom to top  $\Dis  \,\left(=\frac{n_q(2\pi\alpha')}{{\cal N}r_H^3}\right)= 0.001,0.005,0.05, 0.2, 0.4$ and $0.8$. 
}
\end{center}
\end{figure}
 In fig. \ref{twoplots}  we show  plots  for  $\chi$ and $\sigma$ from (\ref{suscepbaryon}) and (\ref{sigma}),  for different values of the baryon density, as a function of the quark mass, $m$.
In the large $m$ limit  $\sigma$ approaches an $n_q$ dependent constant value and $\chi$
dies off as $\sim n_q/m$.
\begin{figure}[ht]
\begin{center}
\includegraphics[scale=0.55]{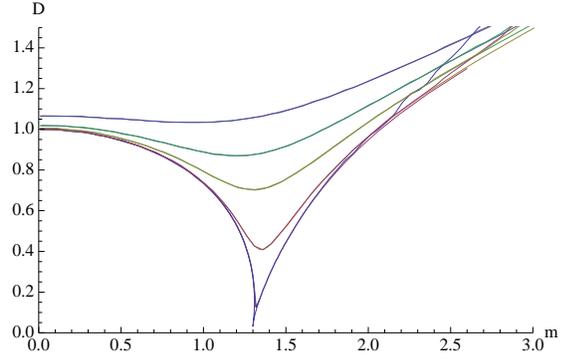}
\caption{\em \label{comparison}
Comparison of $\Diff$ and $\sigma/\chi$ (in units of $1/2\pi T$) against the quark mass defined by the D7 embedding. From bottom to top  $\Dis \,\left(=\frac{n_q(2\pi\alpha')}{{\cal N}r_H^3}\right)= 0.005,0.05, 0.2, 0.4$ and $0.8$. 
}
\end{center}
\end{figure}

\section{The Einstein relation}
For zero baryon number, the analysis performed in \cite{Myers:2007we} confirmed the
Einstein relation
\be
D  = \frac{\sigma}{\chi} \label{einstrel}\, .
\ee
For $n_q\ne 0$  and massless quarks,  the diffusion 
constant simplifies greatly and Einstein;s relation is verified   analytically from (\ref{sigma}), (\ref{suscepbaryon}) and (\ref{diffbaryon}).

For massive quarks, the validity of (\ref{einstrel}) must be established numerically. $\Diff$ can be calculated both from a numerical integration of the coupled equations of motion for the longitudinal fluctuations, or  by integrating 
$\Psi(r)$ and using  (\ref{diffbaryon}). These methods give perfect agreement, pointing to the validity of the hydrodynamic analysis that led to (\ref{diffbaryon}). In  fig. \ref{comparison} we plot the left and right hand sides of (\ref{einstrel}). For all values of $n_q$ both curves agree up to  values of $\psi(r_H)$ close to 1, where numerics become unstable.

\section{Conclusions}
In this note we have rederived and generalized the closed formula obtained in  \cite{Starinets:2008fb} for the diffusion constant $\Diff$ to include the case of a finite baryon density. We have also worked out  the conductivity $\sigma$ and the susceptibility $\chi$ in a holographic setup. We show numerically that for the D3/D7 system,  the three constants obey the Einstein relation (\ref{einstrel}) at finite baryon density.  Formassless quarks this relation holds true in general. 


\end{document}